# Manipulating Multiple Order Parameters via Oxygen Vacancies: The case of $Eu_{0.5}Ba_{0.5}TiO_{3-\delta}$

Weiwei Li[1,2,†], Qian He[3,†], Le Wang[4], Huizhong Zeng[5], John Bowlan[6], Langsheng Ling[7], Dmitry A. Yarotski[6], Wenrui Zhang[8], Run Zhao[1], Jiahong Dai[9], Junxing Gu[4], Shipeng Shen[4], Haizhong Guo[4], Li Pi[7], Haiyan Wang[8], Yongqiang Wang[10], Ivan A. Velasco-Davalos[11], Yangjiang Wu[12], Zhijun Hu[12], Bin Chen[13], Run-Wei Li[13], Young Sun[4], Kuijuan Jin[4], Yuheng Zhang[7], Hou-Tong Chen[6], Sheng Ju[9,*], Andreas Ruediger[11], Daning Shi[1], Albina Y. Borisevich[3,*] and Hao Yang[1,*]

[1]*College of Science, Nanjing University of Aeronautics and Astronautics, Nanjing 211106, China*
[2]*Department of Materials Science and Metallurgy, University of Cambridge, Cambridge CB3 0FS, United Kingdom*
[3]*Materials Science and Technology Division, Oak Ridge National Laboratory, Oak Ridge, Tennessee 37831, USA*
[4]*Beijing National Laboratory for Condensed Matter Physics and Institute of Physics, Chinese Academy of Science, Beijing 100190, China*
[5]*State Key Laboratory of Electronic Thin Films and Integrated Devices, University of Electronic Science and Technology, Chengdu 610054, China*
[6]*Center for Integrated Nanotechnologies, MS K771, Los Alamos National Laboratory, Los Alamos, New Mexico 87545, USA*
[7]*High Magnetic Field Laboratory, Chinese Academy of Science, Hefei 230031, China*
[8]*Materials Science and Engineering Program, Department of Electrical and Computer Engineering, Texas A&M University, College Station, Texas 77843-3128, USA*
[9]*College of Physics, Optoelectronics and Energy, Soochow University, Suzhou 215006, China*
[10]*Materials Science and Technology Division, Los Alamos National Laboratory, Los Alamos, New Mexico 87545, USA*
[11]*Institut National de la Recherche Scientifique — Énergie, Matériaux et Télécommunication (INRS-EMT), 1650 Boul. Lionel Boulet, Varennes J3X 1S2 QC, Canada*
[12]*Center for Soft Condensed Matter Physics and Interdisciplinary Research, Soochow University, Suzhou 215006, China*
[13]*Key Laboratory of Magnetic Materials and Devices, Ningbo Institute of Materials Technology and Engineering, Chinese Academy of Science, Ningbo 315201, China*

Controlling functionalities, such as magnetism or ferroelectricity, by means of oxygen vacancies ($V_O$) is a key issue for the future development of transition metal oxides. Progress in this field is currently addressed through $V_O$ variations and their impact on mainly one order parameter. Here we reveal a new mechanism for tuning both magnetism and ferroelectricity simultaneously by using $V_O$. Combined experimental and density-functional theory studies of $Eu_{0.5}Ba_{0.5}TiO_{3-\delta}$, we demonstrate that oxygen vacancies create $Ti^{3+}$ $3d^1$ defect states, mediating the ferromagnetic coupling between the localized Eu $4f^7$ spins, and increase an off-center displacement of Ti ions, enhancing the ferroelectric Curie temperature. The dual function of Ti sites also promises a magnetoelectric coupling in the $Eu_{0.5}Ba_{0.5}TiO_{3-\delta}$.

Transition metal oxides (TMOs) are attracting significant attention due to their astonishing variety of technologically important physical properties, such as two-dimensional electron gas (2DEG), colossal magnetoresistance (CMR), and multiferroic behavior, etc [1-3]. Tuning the concentration and distribution of ions and vacancies in TMOs provides a route to create and control new functionalities [4]. For many applications, for better or worse, the functionality of TMOs and thin film devices is strongly affected by the formation and distribution of oxygen vacancies ($V_O$). For instance, the introduction of $V_O$ causes a displacement of the Fe ions in $(LaFeO_3)_2/(SrFeO_3)$ superlattices, which induces the polar order [5]. $V_O$ also enable room-temperature ferroelectricity in $SrTiO_3$ thin films by manipulating the $TiO_6$ octahedral tilting around the vacancy site [6]. The electronic properties of these TMOs, especially $ABO_3$-pervoskite structure, are extremely sensitive to structural distortions consisting of cation displacements, deformations, and rotations in an ideal three-dimensional framework of corner-connected $BO_6$ octahedra [7,8]. On the other hand, $V_O$ are well known to play a pivotal role in magnetic properties. Biškup *et al.* suggested that ordered $V_O$ are responsible for insulating ferromagnetism in strained epitaxial $LaCoO_{3-\delta}$ films [9]. Similarly, magnetic phenomena were observed at the $SrTiO_3/LaAlO_3$ interface [10,11] and oxygen-deficient bulk $SrTiO_{3-\delta}$ crystals [12].

Previous studies have shown that it is possible to manipulate the functionality of TMO materials by controlling one order parameter at a time through the concentration or spatial distribution of $V_O$. A natural question arises whether a single experimental parameter, $V_O$, has the ability to simultaneously control multiple order parameters, such as both magnetism and ferroelectricity. In particular, multiferroics with ferromagnetic-ferroelectric (FM-FE) coupling are highly promising for fundamental research and practical applications [13-15]. They are scarce, however, due to the near-incompatibility of the formation of magnetic order (partial filled $d$-orbitals in $3d$ TMOs) and the conventional off-centering mechanism of ferroelectricity (empty $d$-orbitals in $3d$ TMOs) within a single phase [16]. Takahiro *et al* demonstrated theoretically that atomic-size multiferroics emerges in nonmagnetic ferroelectric PbTiO$_3$ through $V_O$ formed at surfaces [17]. While, there are few experimental reports about $V_O$ manipulating magnetism and ferroelectricity in the thin films simultaneously. On the other hand, one can engineer multiferroic properties in ABO$_3$ oxides by chemically controlling the functionality on a site-by-site basis, such as A-site cations providing ferroelectricity and B-site cations supplying magnetism or vice verse. Well known, BiFeO$_3$ (BFO) is the case that ferroelectricity is originated from Bi$^{3+}$ $6s^2$ lone-pair electrons hybridized with O$^{2-}$ $2p^6$ at A-site and antiferromagnetism is derived from Fe$^{3+}$ $3d^5$ at B-site [18]. Unfortunately, the calculations demonstrated that $V_O$ cannot significantly affect the electric polarization, but can slightly alter the value of the macroscopic magnetization of the BFO [19]. The ionic displacements are insensitive to $V_O$, which is responsible for the unaffected electric polarization.

In this letter, we report a new pathway towards a realization of manipulating magnetism and ferroelectricity simultaneously by using $V_O$. Based on previous reports, the criterions that a material must satisfy for this proposed mechanism are as follow: (1) the magnetic and electric ordering should originate from different cations, (2) the ionic displacement should be sensitive to $V_O$. In bulk, $Eu_{0.5}Ba_{0.5}TiO_3$ (EBTO) with a typical $ABO_3$-perovskite structure shows antiferromagnetic (AFM, $T_N \sim 1.9$ K) and ferroelectric (FE, $T_C \sim 213$ K) [20,21]. The AFM and FE are stemmed from the $Eu^{2+}$ $4f^7$ unpaired electrons at A-site and the off-center $Ti^{4+}$ $3d^0$ at B-site, respectively. Moreover, EBTO is structurally similar to the archetypal TMOs, such as $BaTiO_3$ (BTO) and $SrTiO_3$ (STO), and the introduction of $V_O$ has been shown to enhance the ferroelectricity of STO [6,22]. Additionally, our previous results established that the doping of $V_O$ shows strong influence on the magnetic ordering of the $Eu_{0.5}Ba_{0.5}TiO_{3-\delta}$ ($EBTO_{3-\delta}$) thin films [23].

Our present work shows that careful manipulation of $V_O$ can improve both magnetic and FE properties in $EBTO_{3-\delta}$. We experimentally observed that the ferroic orders in $EBTO_{3-\delta}$ thin films are transformed from AFM-FE to FM-FE, and the FE Curie temperature is enhanced to be over room temperature. A small magnetodielectric response was also detected in the $V_O$ doped film, revealing the existence of magnetoelectric coupling. First-principle calculations revealed that the introduction of $V_O$ induces defect associated effects including spin-polarized $Ti^{3+}$ ions, mediating a FM coupling between the local $Eu^{2+}$ $4f^7$ spins, and an enhanced off-center displacement of Ti ions, stabilizing the ferroelectric phase and thus increasing the Curie temperature. The tuning of magnetism and ferroelectricity is both through the medium of Ti sites, which is the origin of the magnetoelectric coupling in $EBTO_{3-\delta}$.

Pulsed laser deposition was used to fabricate $EBTO_{3-\delta}$ films on (001) $SrTiO_3$ (STO) and (001) Nb-doped $SrTiO_3$ (Nb-STO, Nb: 0.5wt%) substrates. All of the $EBTO_{3-\delta}$ films were grown under identical deposition conditions, except for the oxygen pressure, which varied from $1\times10^{-1}$ to $1\times10^{-4}$ Pa (see Supplemental Material for more details [24]). Four kinds of EBTO$_{3-\delta}$ films with different content of $V_O$, grown at oxygen pressure of $1\times10^{-1}$, $1\times10^{-2}$, $1\times10^{-3}$, and $1\times10^{-4}$ Pa, were named as Sample A, B, C, and D, respectively. Moreover, X-ray reciprocal space maps were measured to confirm that strain created by the lattice mismatch of EBTO$_{3-\delta}$ and Nb-STO is fully relaxed (not shown).

To quantitatively determine the stoichiometry and oxygen concentration of the EBTO$_{3-\delta}$ films, we used nuclear resonance backscattering spectrometry (NRBS). The cation ratio in the EBTO$_{3-\delta}$ films (Eu: Ba: Ti) was revealed to be 1 :1 :2. According to the concentrations of the cations and O, the atomicity of O is estimated to be 2.98, 2.96, 2.91, and 2.85 for Samples A, B, C, and D, respectively (see Fig. S1 of Supplemental Material [24]). By comparing the ideal and real atomicity of O, the content of $V_O$ ($\delta$) is calculated to be 0.02, 0.04, 0.09, and 0.15 in Samples A, B, C, and D, respectively. X-ray photoemission spectroscopy (XPS) was used in consideration of very sensitive to variations in the valence state of transition metal ions. The Eu $4d$ and Ba $3d$ spectra exhibit typical Eu$^{2+}$ and Ba$^{2+}$ features, while both Ti$^{3+}$ and Ti$^{4+}$ are observed in Ti $2p$ spectra (see Fig. S2 of Supplemental Material [24]). It is straightforward that Ti$^{3+}$ has one electron at $3d$ orbital (Ti$^{3+}$: $1s^2\ 2s^2\ 2p^6\ 3s^2\ 3p^6\ 4s^0\ 3d^1$), indicating the appearance of Ti$^{3+}$ $3d^1$ states in the EBTO$_{3-\delta}$ films. The presence of the Ti$^{3+}$ $3d^1$ state is consistent with the density-functional theory (DFT) calculations and is believed to have contributed to the FM ordering in the EBTO$_{3-\delta}$ films (see below).

Figure 1(a) and 1(b) show the magnetization versus magnetic field for Samples C and D, respectively. Similar results have also been obtained for Samples A and B (see Fig. S3 of Supplemental Material [24]). Pronounced hysteretic loops are observed, consistent with ferromagnetism, having coercivity of 75.3 and 73.5 Oe for Samples C and D, respectively. Note that the derivative of the magnetization shown as insets has a minimum at around 1.85 K, identified as the FM Curie temperature ($T_C$). In addition, the field dependent magnetization curves are also measured at a higher magnetic field and temperature of 1, 1.5, and 5 K (see Fig. S3 of Supplemental Material [24]). The saturation magnetization, obtained at 1 K, is about 6.72 and 6.80 $\mu_B$/Eu for Samples C and D, respectively, which is close to the ideal magnetic moment of $Eu^{2+}$ ions (7$\mu_B$/Eu).

To further understand $V_O$ effects on magnetic properties, the $V_O$ dependence of coercivity and saturation magnetization are shown in Fig. 1(c). Assuming the local anisotropy energy of ferromagnetism doesn't change significantly with varying the concentration of $V_O$ and according to the Zeeman energy being equal to the anisotropy energy, $E_a = H_c M_s$, the coercivity ($H_C$) gradually decreases with increasing saturation magnetization from Samples A to D. These results demonstrate that the EBTO$_{3-\delta}$ films become ferromagnetism at low temperatures, in contrast to the antiferromagnetism of bulk EBTO. In addition, there is a possibility that the EBTO$_{3-\delta}$ films with even less $V_O$ are also showing ferromagnetism.

To investigate the $V_O$ effects on ferroelectric properties of EBTO$_{3-\delta}$ films, we performed the temperature-dependent optical second harmonic generation (SHG). Optical SHG signals are plotted versus temperature for four samples in Fig. 2(a). Clearly, from Samples A to D, the transition temperature increases from 260 to 395 K, which is significantly larger than that of bulk EBTO (~ 213 K). To further confirm the huge enhancement of the FE $T_C$, we also attempted to measure temperature-dependent dielectric permittivity (see Fig. S4 of Supplemental Material [24]). The curves distinctly show a shift of the maximum in the permittivity (FE $T_C$) from around 255 K for Sample A to 435 K for Sample D. The trend is consistent with SHG results (see Fig. S5 of Supplemental Material [24]), reflecting that the increase in the content of $V_O$ enhances the FE $T_C$ of EBTO$_{3-\delta}$ films. Due to the introduction of $V_O$, the peak in permittivity clearly exhibits a frequency dispersion, which is probably a huge influence of Maxwell-Wagner relaxation derived from the leakage current.

Ferroelectric hysteresis loops were also recorded (see insets of Fig. S4 of Supplemental Material [24]), confirming the ferroelectricity of EBTO$_{3-\delta}$ films. The value of saturated polarization at 150 K is about 14 $\mu$C cm$^{-2}$, which is almost twice of that of bulk EBTO (~8 $\mu$C cm$^{-2}$ at 135 K) [20]. Additionally, the amplitude and phase images of the piezoelectric response measured at 300 K for Sample D were acquired [Fig. 2(b)]. Stable ferroelectric domains with opposite polarization can be written by applying a dc bias to the AFM tip, suggesting room-temperature ferroelectricity and robust polarization. Similar results have also been observed in Samples A to C (see Fig. S4 of Supplemental Material [24]). Moreover, room-temperature piezoresponse hysteresis loops (PHLs) were also obtained and shown in Fig. 2(c). Almost 180° phase contrast is observed in the phase-voltage PHLs, indicating polarization switching. Associated with phase reversal, butterfly shaped amplitude-voltage loops are also observed. The combination of these results proves that the oxygen-deficient EBTO$_{3-\delta}$ films preserve ferroelectricity. Remarkably, the FE $T_C$ was enhanced to be above room temperature, which makes EBTO$_{3-\delta}$ films attractive for the practical applications [25].

Considering the similarity of lattice structure between EBTO and BTO, the ferroelectricity in EBTO is believed to derive from the off-center displacement of Ti ions [20,21,26]. To further confirm the origin of room temperature ferroelectricity in the EBTO$_{3-\delta}$ films, aberration corrected scanning transmission electron microscopy (STEM) measurements were conducted to analyze the off-center displacement of Ti ions in the EBTO$_{3-\delta}$ films. High angle annular dark field (HAADF) imaging in STEM, also known as Z-contrast imaging [27], can be used to precisely measure cation column locations, from which local cation displacement (related to polarization) can be mapped out unit cell by unit cell [28]. The STEM results for Sample D are shown as Fig. 3. An overview of the EBTO$_{3-\delta}$ film is shown in Fig. 3(a), indicating that the film has consistent thickness and uniform appearance on this scale. Close-up looks reveal that some defects have developed in the film. Figure 3(b) shows a medium angle annular dark field (MAADF) image of the region highlighted in Fig. 3(a), in which bright contrast can be seen in the film and at the interface. Since MAADF is sensitive to small lattice distortions [29], such contrast could be from grains in the specimen thickness direction along the electron beam, which are slightly misoriented with each other due to presence of defects such as dislocations. In order to reliably measure displacements of Ti ions, HAADF images were taken in the areas away from those defective areas, where no MAADF contrast can be seen. The cation column positions, determined using a center-of-mass refinement method, were used to calculate the displacements [Fig. 3(c)]. The HAADF image and the resultant Ti ions displacement map for the EBTO$_{3-\delta}$ film and the STO substrate are shown in Fig. 3(d)-3(f) and 3(g)-3(i), respectively. From the displacement maps, it can be seen that the EBTO$_{3-\delta}$ film has non-zero Ti ions displacements in the in-plane ($d_x$) and the out-of-plane ($d_y$) directions. While the absolute value of the displacements is fairly small and approaches the detection limit for the technique, the histogram shown in Fig. 3(j) shows unambiguously that the average value of $d_x$ (blue) and $d_y$ (black) for the EBTO$_{3-\delta}$ film is distinct from zero, namely about 0.07 Å and 0.03 Å, respectively. This finding is consistent with the SHG and PFM results confirming that the FE $T_C$ of Sample D is above room temperature. In contrast to that, the average value of $d_x$ (red) and $d_y$ (green) for the STO substrate (calculated the same way) is about zero, which is consistent with its room temperature paraelectricity.

To understand the physical process underlying the manipulation of multiple order parameters in the EBTO$_{3-\delta}$ films, DFT calculations were performed. A-type atomic arrangement of Eu and Ba ions was used in the calculations due to the simultaneous lowest energy and AFM-FE (see Fig. S6 of Supplemental Material [24]). To further shed light on $V_O$ effects, electron distribution under different configurations of $V_O$ position (see Fig. S7 of Supplemental Material [24]) is investigated and shown in Fig.

4(a). The change of electron distribution around Ti sites is clearly observed upon the presence of $V_O$, indicating the appearance of $Ti^{3+}$ $3d^1$ states. While, the valence states of Eu and Ba ions remain divalent. Differential charge distribution between AFM and FM orders is also shown in Fig. 4(b). The electron around oxygen and Eu sites show spatially asymmetric variation. In particular, when $V_O$ is located at $TiO_2$ plane, the electron distribution around Eu sites shows obvious differences between each other, implying a hybridization of $Eu^{2+}$ $4f^7$ and $Ti^{3+}$ $3d^1$. Note that the FM states of all $V_O$ configurations are energetically more favorable than their AFM states [Fig. 4(b)].

Based on the results given above, we now focus on understanding the effects of $V_O$ on FM and FE orders by presenting a model and a band diagram [Figure 4(c) and 4(d), respectively]. Before taking into account $V_O$, superexchange coupling between $Eu^{2+}$ $4f^7$ spins via $Ti^{4+}$ $3d^0$ states and off-center displacement of Ti ions are responsible for AFM and FE orders observed in bulk EBTO [20,21,30], respectively [Fig. 4(c)]. Combined with XPS valence band spectra [30,31], the existence of $V_O$ creates $Ti^{3+}$ $3d^1$ defect states, localizing within the band gap and overlapping with $Eu^{2+}$ $4f^7$ states [Figure 4(d)]. In this case, the spin-polarized $Ti^{3+}$ will mediate FM coupling between the localized $Eu^{2+}$ $4f^7$ spins in $EBTO_{3-\delta}$ [Fig. 4(c)]. Furthermore, at the presence of $V_O$, Ti ions with remaining oxygen form pyramid structure instead of oxygen octahedra (see Fig. S7 of Supplemental Material [24]) and increase the $d_{3z^2-r^2}$ or $d_{xy}$ character of local orbitals of $Ti^{3+}$ ions adjacent to the $V_O$ sites [32,33]. When $V_O$ is situated in the EuO or BaO plane, Ti ions move naturally towards $V_O$ to avoid electrostatic interaction and the $d_{3z^2-r^2}$ occupation can lead to a local polar distortion. On the other hand, when $V_O$ is placed in the TiO$_2$ plane, $d_{xy}$ orbital is preferred, resulting in an additional polar distortion in the TiO$_2$ plane. These local distortions should couple with globe polar distortion in pristine EBTO and will afford a totally new degree of freedom to tune the ferroelectricity in EBTO$_{3-\delta}$. In other words, the off-center displacement of Ti ions will be enhanced by the introduction of $V_O$, thereby enhancing the FE Curie temperature in EBTO$_{3-\delta}$ [Fig. 4(c)]. These results definitely approve that tuning $V_O$ can effectively change magnetic and electric degrees of freedom in EBTO$_{3-\delta}$ simultaneously. More than this, it should be emphasized that the manipulating of magnetism and ferroelectricity is both through the medium of Ti sites, revealing the existence of magnetoelectric coupling in EBTO$_{3-\delta}$. The coupling between electric and magnetic orders was confirmed in the $V_O$ doped film by the magnetodielectric measurements [Fig. 5]. Due to the spin-phonon coupling, as shown in Fig. 5(a), the dielectric constant shows a dependence on the external magnetic fields in the FM-FE state [34-36]. In contrast, as shown in Fig. 5(b), the influence of magnetic field is almost negligible in the PM-FE state.

In conclusion, a new mechanism is proposed for controlling multiple order parameters simultaneously by using a single experimental parameter, $V_O$. EBTO$_{3-\delta}$ was chosen to realize this strategy because magnetism and ferroelectricity are originated from different cations and the off-center displacements of Ti ions are sensitive to $V_O$. The emergence of ferromagnetism is the result of oxygen vacancy-created Ti$^{3+}$ 3$d^1$ defect states, mediating ferromagnetic coupling between the localized Eu 4$f^7$ spins. On the other hand, the introduction of $V_O$ increases an off-center displacement of Ti ions, enhancing the ferroelectric Curie temperature of EBTO$_{3-\delta}$. The dual function of Ti sites induces magnetoelectric coupling, which reinforces the high potential of oxygen vacancies engineering as a tool for designing oxide thin films suitable for multifunctional device applications.

The authors thank Kelvin H. L. Zhang for valuable discussion, and also acknowledge the support of the National Basic Research Program of China (No. 2014CB921001), the National Natural Science Foundation of China (Grant No. 11274237, U1632122, 11004145, 51202153, U1332209, U1435208, 11134012, 11174355, 11474349, and 11227405), and the Program for Postgraduates Research Innovation in University of Jiangsu Province under No. CXZZ13_0798. The STEM studies (QH and AYB) is supported by the U.S. Department of Energy, Office of Science, Basic Energy Sciences, Materials Sciences and Engineering Division. The TEM studies at Texas A&M University is funded by the U.S. National Science Foundation (DMR-1643911 and DMR-1565822). Ion beam analysis (YW) and SHG measurements are supported by the Center for Integrated Nanothechnologies (CINT), a US DOE Nanoscale Research Center, jointly operated by Los Alamos and Sandia National laboratories. AR gratefully acknowledges financial support from NSERC through a discovery grant, from FRQNT and from CFI through the leaders opportunity fund.

†W. Li and Q. He contributed equally to this work.

*Corresponding authors:
 yanghao@nuaa.edu.cn; albinab@ornl.gov; jusheng@suda.edu.cn

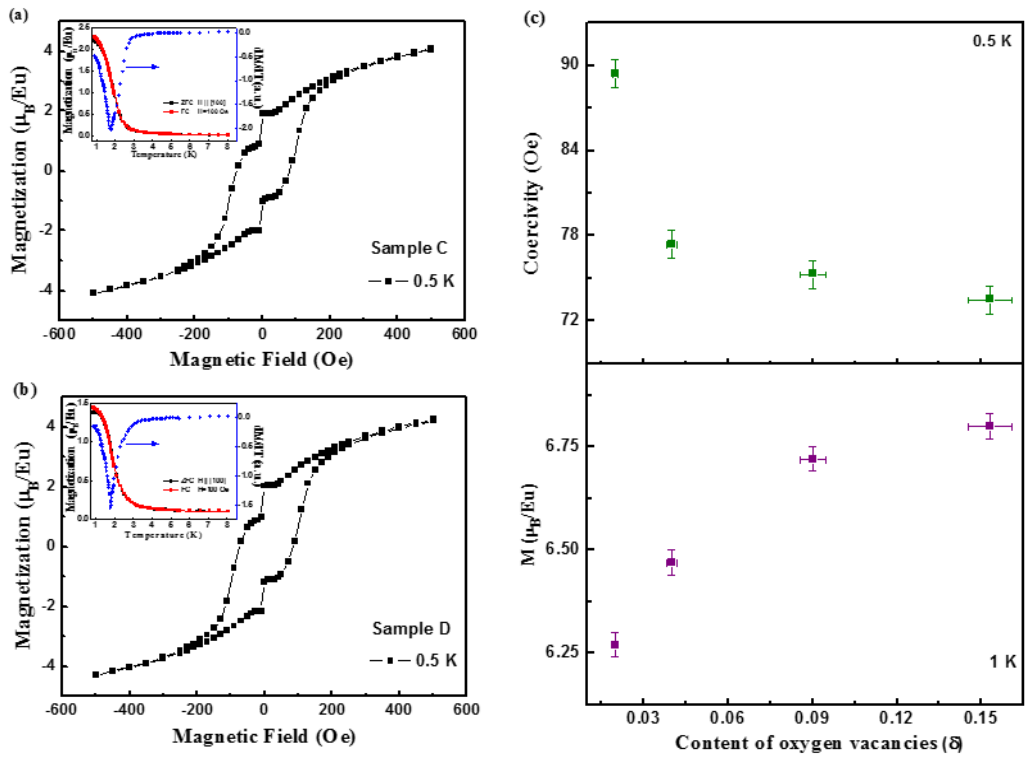

Figure 1 Magnetic hysteresis loops for Sample (a) C and (b) D. Insets show temperature dependence of magnetization curves and the derivative of magnetization with respect to the temperature (obtained from FC curves). (c) The content of $V_O$ ($\delta$) dependences of coercivity and saturation magnetization.

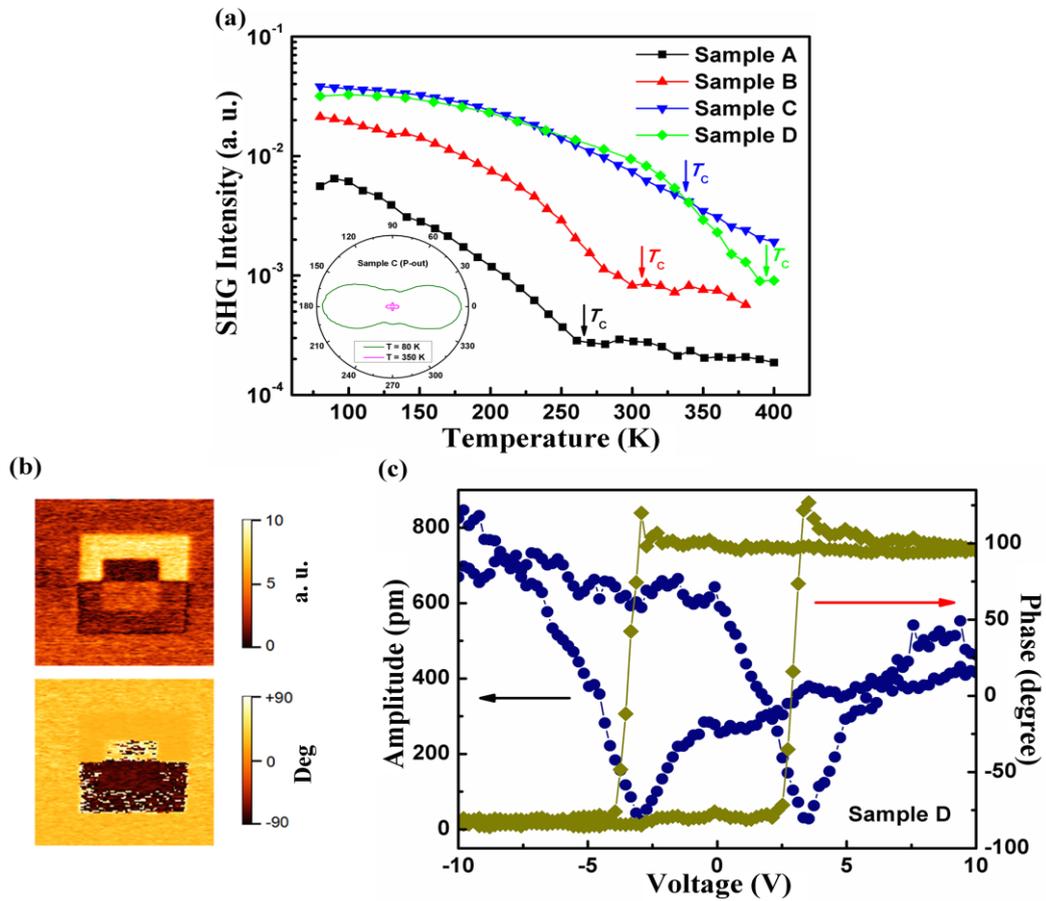

Figure 2 (a) SHG intensity corresponding to the 'P' component of SHG for 'P' polarized fundamental as a function of temperature for Samples A to D. The inset shows polar plot of SHG intensity (radius) versus fundamental polarization (azimuthal angle) at 80 and 350 K for 'P' for Sample C. (b) The PFM amplitude (upper panel) and phase (lower panel) images of the rectangular ferroelectric domain patterns written by a biased tip in Sample D at 300 K. The scan size is 2 µm. (c) Room-temperature piezoresponse amplitude and phase hysteresis loops of Sample D.

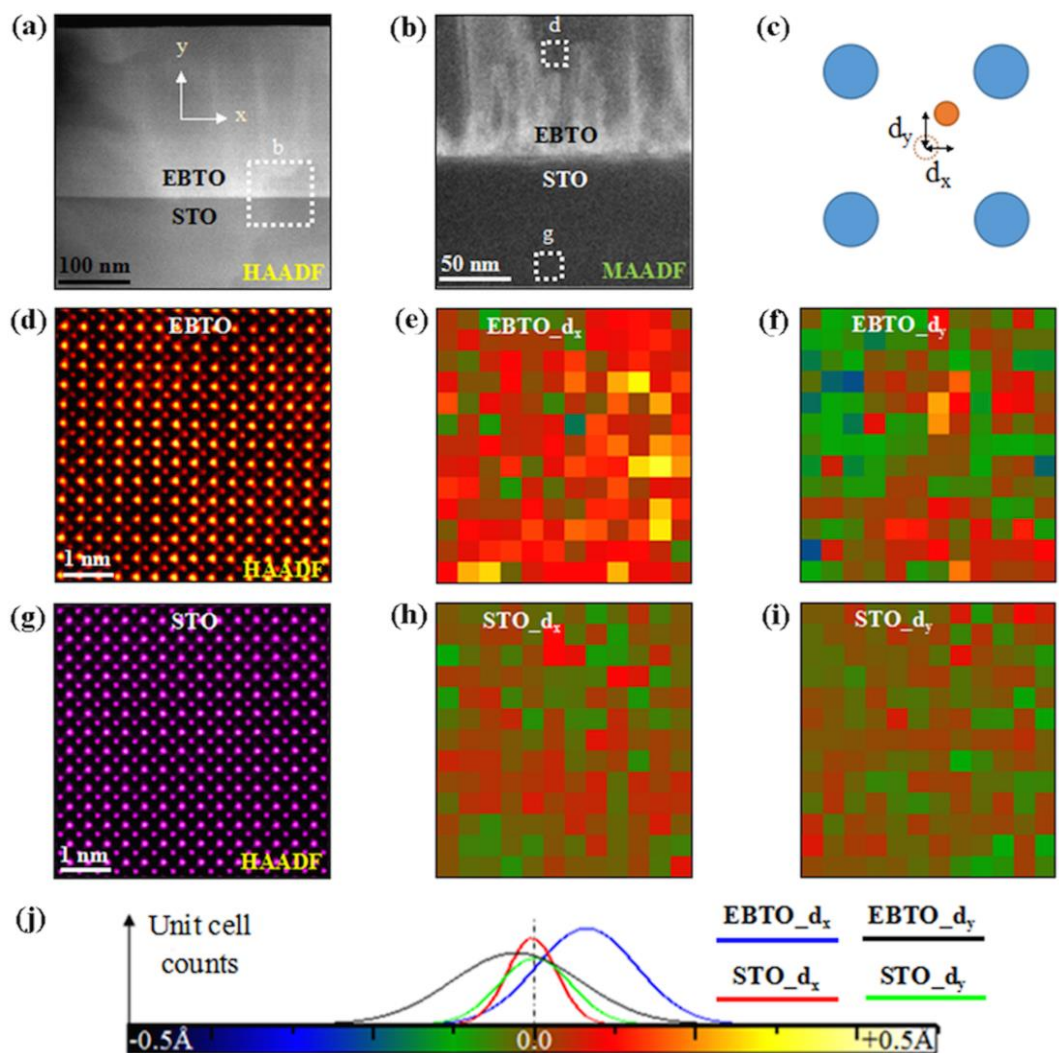

Figure 3 (a) Lower magnification HAADF-STEM image of Sample D. (b) MAADF-STEM image of the highlighted region in (a). (c) Schematic of measuring in-plane ($d_x$) and out-of-plane ($d_y$) displacement of B site cations (orange) from the center position with respect to the A site cations (blue). (d-f) Higher magnification HAADF-STEM image of EBTO region highlighted in (b) and the resultant Ti ion displacement map. (g-i) Higher magnification HAADF-STEM image of STO region highlighted in (b) and the resultant Ti ions displacement map. (j) The statistical histogram of Ti ions displacements in (d) and (g), and the color scheme used in the displacement maps.

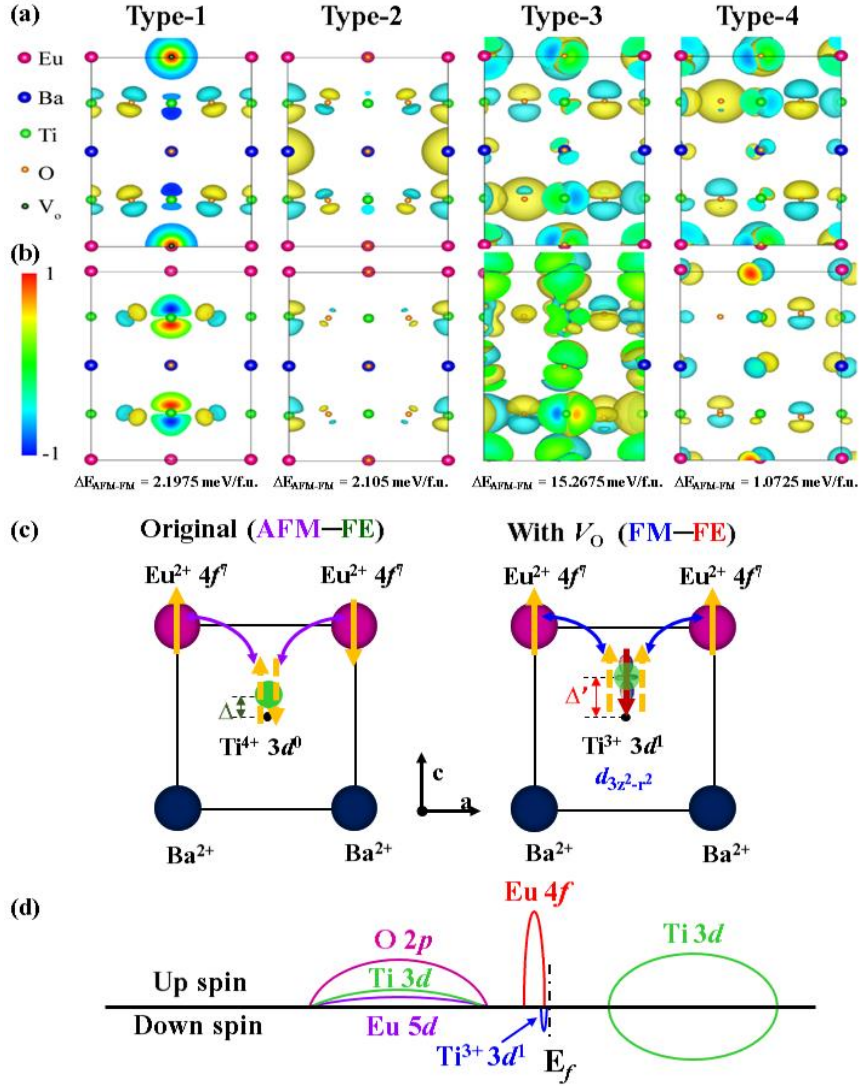

Figure 4 (a) Differential charge upon the presence of $V_O$. (b) Differential charge between AFM order and FM order of EBTO$_{3-1/4}$. Type-1: $V_O$ at the EuO plane; Type-2: $V_O$ at the BaO plane; Type-3 and Type-4: $V_O$ at the TiO$_2$ plane. In all cases, FM ordering is favored with the presence of $V_O$. (c) Sketch of the effects of $V_O$ on ferromagnetism and ferroelectricity in the oxygen-deficient EBTO$_{3-\delta}$. Left panel: Original AFM and FE orders in bulk EBTO. Right panel: FM and FE orders with $V_O$ at the EuO or BaO plane in the EBTO$_{3-\delta}$. (d) Band diagram of the oxygen-deficient EBTO$_{3-\delta}$.

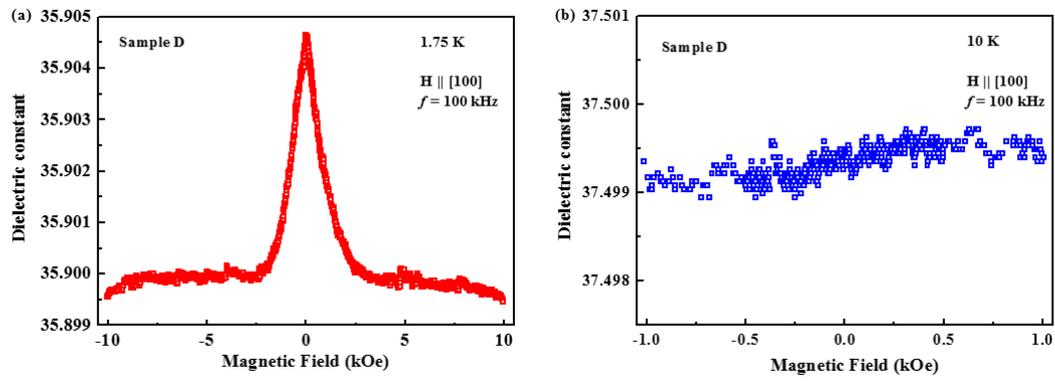

Figure 5 The magnetic field dependence of dielectric constant measured at (a) 1.75 K and (b) 10 K of Sample D.

# Supplemental Material for Manipulating Multiple Order Parameters via Oxygen Vacancies: The case of Eu$_{0.5}$Ba$_{0.5}$TiO$_{3-\delta}$

Weiwei Li[1,2,†], Qian He[3,†], Le Wang[4], Huizhong Zeng[5], John Bowlan[6], Langsheng Ling[7], Dmitry A. Yarotski[6], Wenrui Zhang[8], Run Zhao[1], Jiahong Dai[9], Junxing Gu[4], Shipeng Shen[4], Haizhong Guo[4], Li Pi[7], Haiyan Wang[8], Yongqiang Wang[10], Ivan A. Velasco-Davalos[11], Yangjiang Wu[12], Zhijun Hu[12], Bin Chen[13], Run-Wei Li[13], Young Sun[4], Kuijuan Jin[4], Yuheng Zhang[7], Hou-Tong Chen[6], Sheng Ju[9,*], Andreas Ruediger[11], Daning Shi[1], Albina Y. Borisevich[3,*] and Hao Yang[1,*]

[1]*College of Science, Nanjing University of Aeronautics and Astronautics, Nanjing 211106, China*
[2]*Department of Materials Science and Metallurgy, University of Cambridge, Cambridge CB3 0FS, United Kingdom*
[3]*Materials Science and Technology Division, Oak Ridge National Laboratory, Oak Ridge, Tennessee 37831, USA*
[4]*Beijing National Laboratory for Condensed Matter Physics and Institute of Physics, Chinese Academy of Science, Beijing 100190, China*
[5]*State Key Laboratory of Electronic Thin Films and Integrated Devices, University of Electronic Science and Technology, Chengdu 610054, China*
[6]*Center for Integrated Nanotechnologies, MS K771, Los Alamos National Laboratory, Los Alamos, New Mexico 87545, USA*
[7]*High Magnetic Field Laboratory, Chinese Academy of Science, Hefei 230031, China*
[8]*Materials Science and Engineering Program, Department of Electrical and Computer Engineering, Texas A&M University, College Station, Texas 77843-3128, USA*
[9]*College of Physics, Optoelectronics and Energy, Soochow University, Suzhou 215006, China*
[10]*Materials Science and Technology Division, Los Alamos National Laboratory, Los Alamos, New Mexico 87545, USA*
[11]*Institut National de la Recherche Scientifique — Énergie, Matériaux et Télécommunication (INRS-EMT), 1650 Boul. Lionel Boulet, Varennes J3X 1S2 QC, Canada*
[12]*Center for Soft Condensed Matter Physics and Interdisciplinary Research, Soochow University, Suzhou 215006, China*
[13]*Key Laboratory of Magnetic Materials and Devices, Ningbo Institute of Materials Technology and Engineering, Chinese Academy of Science, Ningbo 315201, China*

1. **Film fabrication and structure characterization**

Epitaxial Eu$_{0.5}$Ba$_{0.5}$TiO$_{3-\delta}$ (EBTO$_{3-\delta}$) thin films were grown by pulsed laser deposition using a pulsed excimer laser (Lambda Physik, 248 nm, 3 Hz, 2 J/cm$^2$) from a high-density Eu$_{0.5}$Ba$_{0.5}$TiO$_3$ ceramic target. The films were deposited on (001) oriented SrTiO$_3$ (STO) and Nb-doped SrTiO$_3$ (Nb-STO, Nb: 0.5wt%) substrates at a temperature of 700 °C. Oxygen pressure ranged from $1\times10^{-1}$ to $1\times10^{-4}$ Pa was used with the purpose of introducing $V_O$ with variable content. To further increase the content of $V_O$ and to relax the out-of-plane lattice strain, all the films were post-annealed at 1000 °C under a flowing gas of 95 vol% Ar+5 vol% H$_2$ for 10 hours after deposition [1-3].

The crystal structures of EBTO$_{3-\delta}$ thin films were investigated by X-ray diffraction (XRD, Rigaku K/Max) and transmission electron microscopy (TEM, FEI Tecnai F20 analytical microscope). Typical XRD $\theta$-$2\theta$ patterns suggested that the EBTO$_{3-\delta}$ phase is oriented along the *c*-axis. The in-plane orientation of EBTO$_{3-\delta}$ thin films with respect to the major axis of the STO substrate is revealed by phi-scans. The epitaxial relationship is determined to be (002)$_{EBTO}$||(002)$_{STO}$ and [103]$_{EBTO}$||[103]$_{STO}$. The TEM images revealed that the film grows coherently on the STO substrate without any apparent interface reaction and intermixing along the interface, which confirms epitaxy through good lattice match between the film and the substrate. The film thicknesses, revealed by TEM, were 150 to 300 nm. The displacement of Ti ions in the EBTO$_{3-\delta}$ thin films was measured by scanning transmission electron microscopy (STEM) using Nion UltraSTEM 200 in Oak Ridge National Laboratory with aberration correction to 5$^{th}$ order. The cold field emission gun was operated at 200 kV and the beam convergence angle was 30 mrad. The medium angle annular dark field (MAADF) detector has a collection angle of 28 mrad to 80 mrad, the latter of which is the minimum collection angle of the High angle annular dark field (HAADF) detector. The cross-section specimen for STEM observation was prepared by mechanical polishing and ion-milling.

## 2. Cations stoichiometry and oxygen concentration in EBTO$_{3-\delta}$ thin films

To quantitatively determine the stoichiometry and oxygen concentration of the EBTO$_{3-\delta}$ thin films, we used nuclear resonance backscattering spectrometry (NRBS). In principle, NRBS is identical to the well-known traditional Rutherford backscattering spectrometry (RBS) in terms of scattering kinematics or the measured He-energies, which identifies different mass of elements or represents different depths for a given element. However, NRBS has two advantages in this work: its much higher detection sensitivity of oxygen than RBS and its good depth resolution to selectively measure the oxygen in the EBTO$_{3-\delta}$ films on an oxygen-rich substrate, such as STO in the present work [4,5]. In addition, higher energy allows the interaction between the projectile and the target nucleus to overcome the Coulomb repulsive force barrier due to a very close encounter between incident particle and the target nucleus. As a result, the scattering cross section becomes non-Rutherford and often exhibits a narrow and strong resonance peaks that sometimes can be orders of magnitude stronger that the Rutherford value. In our measurement, the beam energy was chosen in a way that resonant scattering occurs in the film (not in the substrate).

NRBS with a He$^+$ beam energy of 3.043 MeV was performed on 3 MV Pelletron Tandem Accelerator at Los Alamos National Laboratory. In NRBS, the backscattering detector with a solid angle of ~ 3.8 msr was mounted at 167º from the beam direction. In the NRBS spectrum, Y-axis represents the measured counts of the backscattering He particles while X-axis represents the measured energies of the backscattered He particles. The channel numbers are the direct registration of the particle energies from the digital data acquisition system. The particle energies could reflect two different things. For a given element (e.g. Ti), different energies represent different depths for that element with lower energies meaning deeper regions below the surface, which allows us to distinguish Ti in the film (higher energy) from Ti in the substrate (lower energy). For a given depth (e.g. at the surface), different energies represent different masses (elements) with lower energies meaning lighter elements.

As a He$^+$ beam energy of 3.043 MeV was used, the He scattering from the heavier elements (Eu, Ba, and Ti) in the films is still Rutherford, but there is a strong and narrow resonant scattering from oxygen. Figure S1(a) shows the RBS spectrum measured from Sample D. Similar results have also been obtained from Samples A, B, and C (not shown). Within the uncertainty of the measurements (~ 5%), the overlaid results of a SIMNRA [6,7] simulated spectrum with the measured RBS spectrum confirmed that the cation ratio in the EBTO$_{3-\delta}$ thin films (Eu: Ba: Ti) is 1 :1 :2.

Figure S1(b) shows the measured oxygen scattering spectrum in Sample D along with a SIMNRA fitted oxygen spectrum. Similar results have also been obtained from Samples A, B, and C (not shown). To minimize uncertainties related to the nuclear scattering cross section data, the scattering angle, the detector solid angle, the beam charge integration, and the incident beam energy, a bare STO substrate as a standard reference was measured and analyzed along with our EBTO$_{3-\delta}$ thin films grown on STO substrates. According to the concentrations of the cations and O, the atomicity of O is estimated to be 2.98, 2.96, 2.91, and 2.85 for Samples A, B, C, and D, respectively.

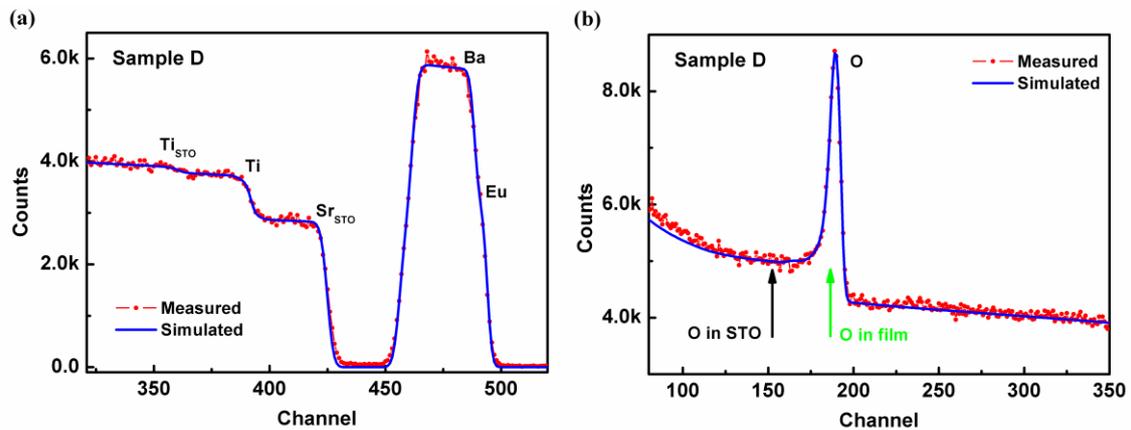

FIG. S1. Stoichiometry and oxygen concentration of Sample D studied by NRBS. (a) RBS spectrum and (b) oxygen resonance peak with the simulated curve (blue line).

## 3. Analysis of valence states of cations in EBTO$_{3-\delta}$ thin films

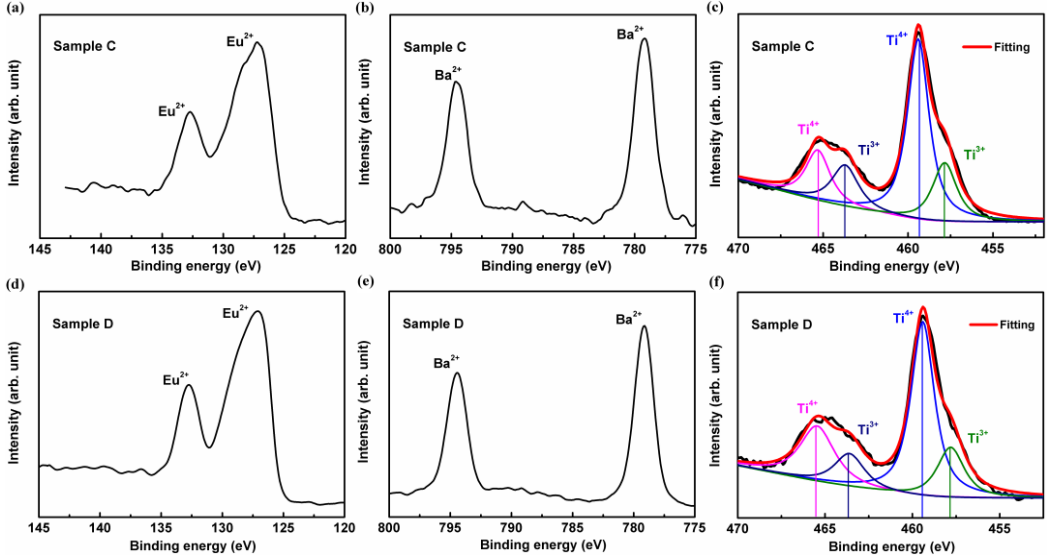

FIG. S2. The valence states of Samples C and D were investigated by X-ray photoemission spectroscopy. Spectra of (a) and (d) Eu 4$d$, (b) and (e) Ba 3$d$, and (c) and (f) Ti 2$p$ peaks. All spectra were aligned to the O 1s at 530.1 eV. In the bulk Eu$_{0.5}$Ba$_{0.5}$TiO$_3$, the valence states of Eu and Ba ions are 2+ and Ti ions are 4+. The presence of charge carriers of Ti 3$d^1$ character must alter the shape of the Ti core levels for effectively corresponding to Ti$^{3+}$ states (navy and olive lines at the binding energy around 463.5 eV and 457.8 eV, respectively), which differ in binding energy from the host Ti$^{4+}$ environment (magenta and blue lines at the binding energy around 459.4 eV and 465.4 eV, respectively). It is straightforward that Ti$^{3+}$ has one electron at 3$d$ orbital (Ti$^{3+}$: 1$s^2$ 2$s^2$ 2$p^6$ 3$s^2$ 3$p^6$ 4$s^0$ 3$d^1$). In other words, Ti$^{3+}$ features show up at the low binding energy side of the Ti 2p core level indicating the presence of Ti$^{3+}$ 3$d^1$ states.

## 4. Study of ferromagnetism in EBTO$_{3-\delta}$ thin films

Magnetic data was acquired using a superconducting quantum interface device magnetometer (SQUID) equipped with a He$^3$ insert (Quantum Design, MPMS-XL). The magnetic field is applied in the plane of the sample with a structure of EBTO$_{3-\delta}$/STO and temperature is down to 0.5 K. The measurements of zero-field-cooled (ZFC) and field-cooled (FC) conditions are performed on an external magnetic field of 100 Oe.

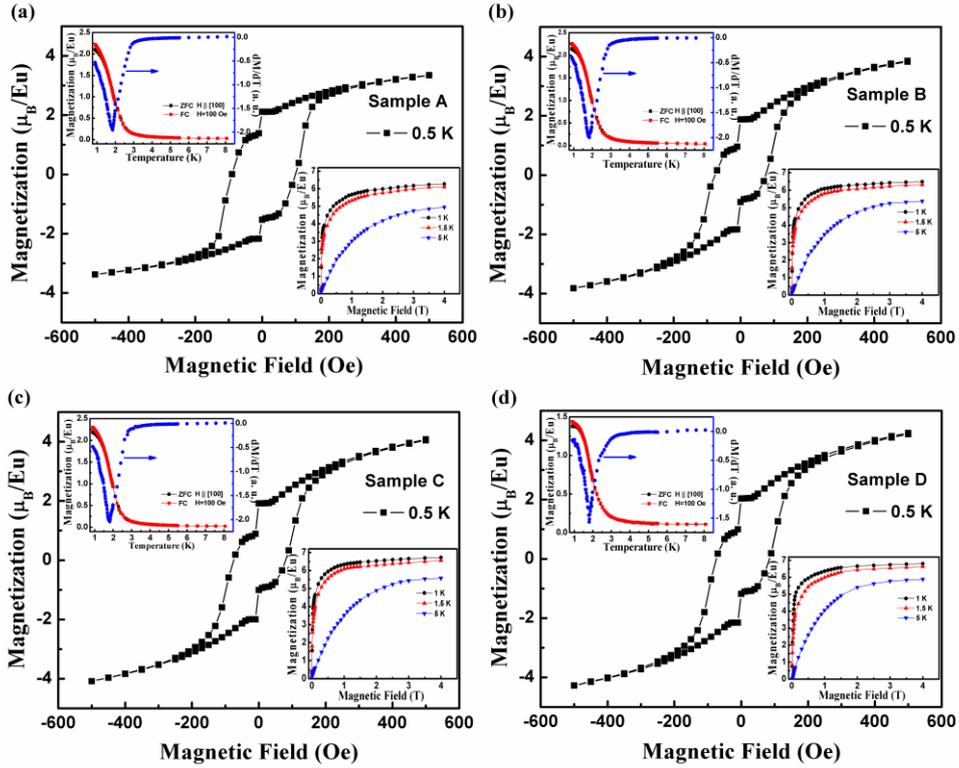

FIG. S3. Magnetic properties of Samples A, B, C, and D. Magnetic hysteresis loops measured at 0.5 K for Samples (a) A, (b) B, (c) C, and (d) D. Top-left insets show temperature dependence of magnetization curves and the derivative of the magnetization with respect to the temperature (obtained from the FC curves). It has been found that the derivative of the magnetization has a minima at around 1.85 K, meaning a ferromagnetic Curie temperature of 1.85 K. Bottom-right insets show magnetic-field dependent magnetization curves. It can be seen that the magnetization increased sharply at low fields and then approached saturation quickly at 1 and 1.5 K, which is also an evidence for the ferromagnetic ordering below 1.85 K. These results

verified that the EBTO$_{3-\delta}$ thin films become ferromagnetic. Furthermore, the coercivity and saturation magnetizations can be manipulated by varying the content of $V_O$.

## 5. Electric properties of EBTO$_{3-\delta}$ thin films

### 5.a. Second harmonic generation

The dominant electric dipole contribution to optical second harmonic generation (SHG) is exhibited only by materials that lack inversion symmetry. Because ferroelectrics lack inversion symmetry, SHG is generally strongly enhanced in ferroelectrics below the ferroelectric Curie temperature. Thus SHG represents good probe for ferroelectricity without applying voltages. Temperature-dependent SHG measurements were performed in reflection with an 800 nm, 100 fs pulsed Ti:Sapphire laser. The angle of incidence was 45 degrees, and the S and P components of the reflected second harmonic were measured as a function of the polarization of the 800 nm fundamental beam. The SHG light was filtered to remove the reflected 800 nm beam, and detected with a photomultiplier using lock-in detection. The average power incident on the sample was 10 mW which corresponds to a fluence of ~ 3.5 mJ/cm$^2$.

### 5.b. Dielectric and ferroelectric characterization

For electrical properties measurements, vertical sandwich capacitors with a structure of Pt/EBTO$_{3-\delta}$/Nb-STO were fabricated, for which the films with a thickness of ~250 nm were used and the Pt top electrodes with an area of $8\times10^{-4}$ cm$^2$ were deposited by sputtering. The dielectric properties were investigated using an Agilent 4294A Impedance Analyzer. The ferroelectric hysteresis loops were measured at 5 kHz using a ferroelectric test system (Radiant RT6000HVS). These measurements were performed at previously indicated temperatures in a Linkam Scientific Instruments HFS600E-PB4 system.

### 5.c. Piezoresponse force microscopy

Temperature-dependent Piezoresponse force microscopy (PFM) was measured using a commercial atomic force microscopy (SPA300-HV, Seiko Instruments) with a lock-in module (HF2LI, Zurich Instruments). During the low temperature PFM measurement, the sample is mounted on a liquid Nitrogen cooling stage in high vacuum (~$10^{-5}$ Pa) and temperature accuracy is kept about 0.5 K. For domain imaging, an a.c. voltage with the peak-to-peak amplitude of 2.5 V and frequency of 74 kHz was applied to the Pt-coated silicon AFM tip. For domain switching, a d.c. bias of +8 V and -8 V were applied to write bright and dark domain patterns (PFM phase images) respectively. Local piezoelectric hysteresis loops were detected using a commercial AIST-NT Smart 1000 system with a Pt-Ir-coated beam deflection cantilever off the contact resonance frequency.

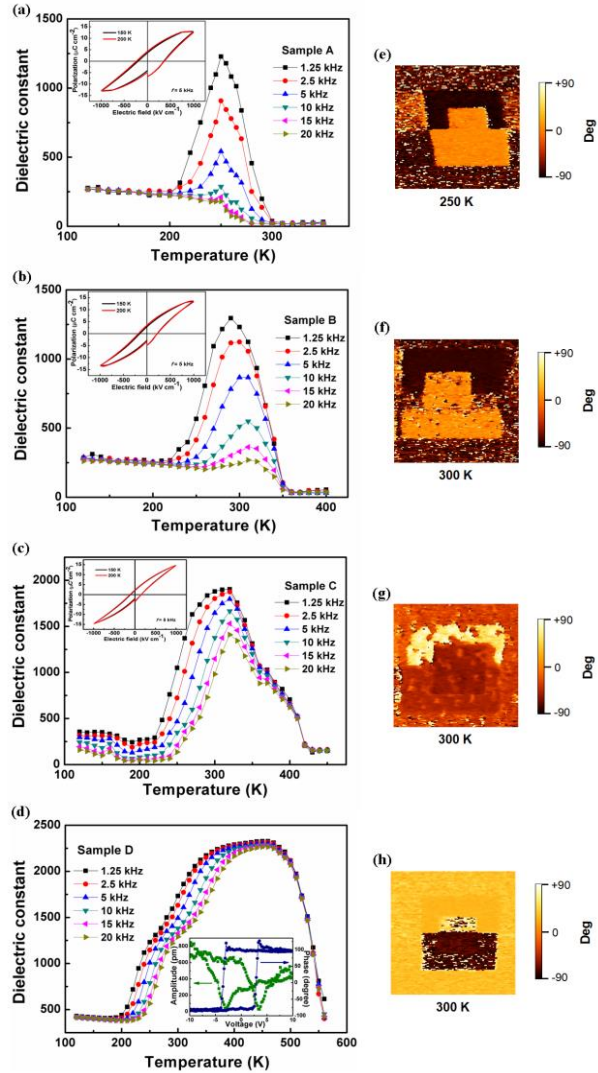

FIG. S4. Electric properties of Samples A, B, C, and D. Temperature dependence of dielectric permittivity for Samples (a) A, (b) B, (c) C, and (d) D measured at various frequencies. Figures (e-h) show the phase images of the pizeoelectric response for Samples A, B, C, and D measured at 250, 300, 300, and 300 K respectively. The scan size is 2 µm. Insets of (a)-(c) show ferroelectric hysteresis loops measured at 150 and 200 K with a frequency of 5 kHz. Inset of (d) shows room-temperature pizeoresponse amplitude and phase hysteresis loops measured by PFM. The combination of these results proves that the EBTO$_{3-\delta}$ thin films preserve ferroelectricity with an enhanced ferroelectric Curie temperature.

**5.d. Comparison of ferroelectric Curie temperature**

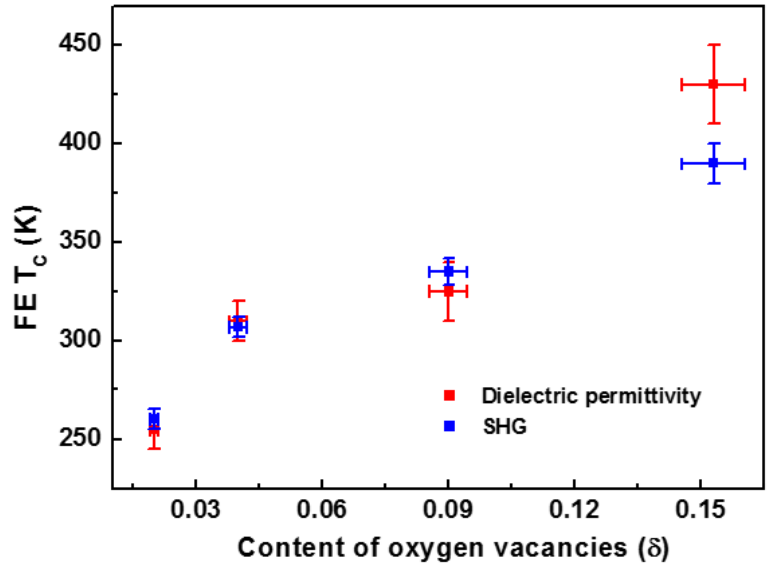

FIG. S5. The content of $V_O$ ($\delta$) dependence of ferroelectric Curie temperature (FE $T_C$) obtained from dielectric permittivity and optical second harmonic generation (SHG) measurements. It can be seen that manipulating the content of $V_O$ gradually increases the ferroelectric Curie temperature from Samples A to D.

## 6. Density-Functional Theory

Our ab initio calculations are performed using the accurate full-potential projector augmented wave (PAW) method [8], as implemented in the Vienna *ab initio* Simulation Package (VASP) [9]. They are based on the density-functional theory (DFT) with the generalized gradient approximation (GGA) in the form proposed by Perdew, Burke, and Ernzerhof (PBE) [10]. The strong Coulomb correlation is included within the GGA+U method and the Hubbard U is 4 eV for the Eu 4*f* orbitals and 3 eV for the Ti 3*d* orbitals [11]. A large plane-wave cutoff of 500 eV is used throughout and the convergence criteria for energy is $10^{-5}$ eV. PAW potentials are used to describe the electron-ion interaction with 17 valence electrons for Eu ($4f^7 5s^2 5p^6 6s^2$), 10 for Ba ($5s^2 5p^6 6s^2$), 10 for Ti ($3p^6 3d^2 4s^2$), and 6 for O ($2s^2 2p^4$). A $8\times8\times6$ Monkhorst-Pack k-point mesh is used in the Brillouin zone integration for the 20 atoms system. For larger systems, the k-points mesh is reduced accordingly. In our calculations, ions are relaxed towards equilibrium positions and lattice constants are optimized with the convergence criteria for energy of $10^{-5}$ eV. For the intrinsic property of $Eu_{0.5}Ba_{0.5}TiO_3$, we also calculate the band structure and enhance band-gap within the framework of hybrid functional [12], whose accuracy approaches the advance GW method [13], but become more computationally feasible in dealing with transition metal oxides.

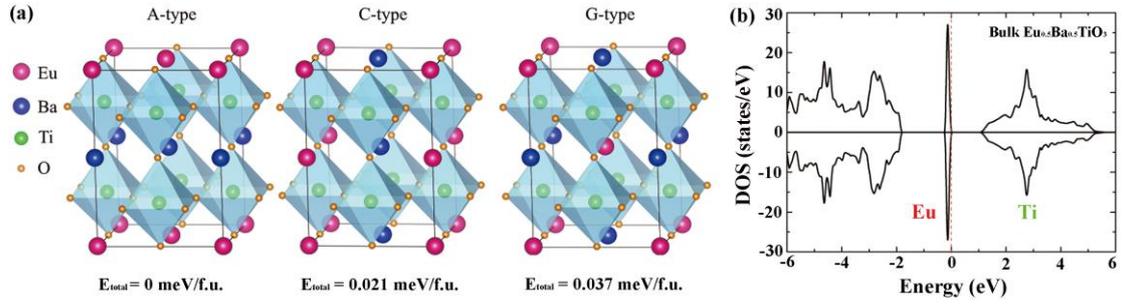

FIG S6. (a) Three types of Eu and Ba atomic orderings in 20-atom unit cell. From left to right is A-type, where EuO and BaO planes stack along the c-axis; C-type, where Eu and Ba ions are distributed along the c-axis; and G-type, where all Eu nearest neighbors at A-site is Ba. Total energy of these three structures listed below show A-type structure is energetically favorable. (b) Total density of states (DOS) for bulk $Eu_{0.5}Ba_{0.5}TiO_3$ with the A-type magnetic configuration. The Fermi level is indicated by the dotted line.

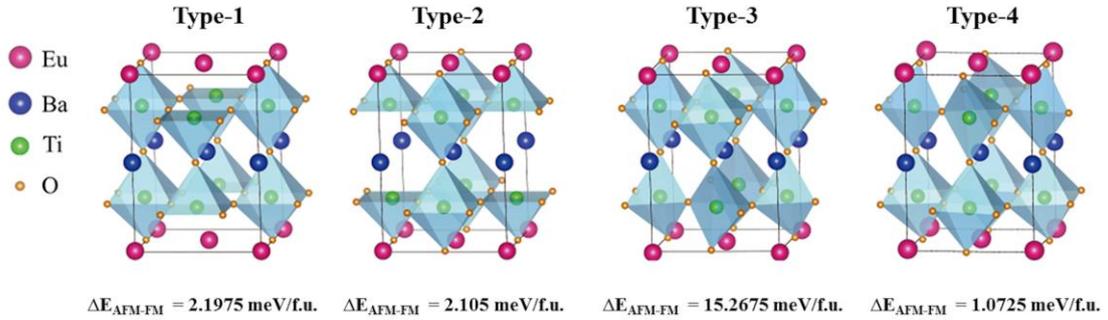

FIG S7. Four types of oxygen vacancies ($V_O$) position in the A-type $Eu_{0.5}Ba_{0.5}TiO_{3-\delta}$. Type 1 with $V_O$ at EuO plane, Type 2 with $V_O$ at BaO plane, and Type 3 and Type 4 with $V_O$ at $TiO_2$ plane. In all cases, comparing energy differences between AFM and FM orders, FM ordering is favored with the presence of $V_O$.

## 7. Coupling of magnetic and electric orders in EBTO$_{3-\delta}$ thin films

Magnetodielectric measurements were used to reveal the coupling between magnetic and electric orders in EBTO$_{3-\delta}$ thin films. Magnetic field (applied in-plane of the sample with a structure of Pt/EBTO$_{3-\delta}$/Nb-STO) in the measurements was controlled by inserting the sample in a Cryogen-free Superconducting Magnet System (Oxford Instruments, TeslatronPT) with a home-made probe.